\documentclass{llncs}

\newcommand{\mod}{\mbox{ mod }}
\newcommand{\Zq}{(Z/qZ)^*}

\newtheorem{deff}{Definition}

\newcommand{\arrowsize}{6cm}

\title{
Pretty-Simple Password-Authenticated Key-Exchange Protocol
}

 \author{Kazukuni Kobara and Hideki  Imai}
         
 \institute{Institute of Industrial Science, The University of Tokyo\\
 4-6-1, Komaba, Meguro-ku, Tokyo, 153-8505 Japan\\
   TEL : +81-3-5452-6232 \\
   FAX : +81-3-3452-6631 \\
   E-mail: \{kobara,imai\}@iis.u-tokyo.ac.jp}
     
\begin{document}
\maketitle

\begin{abstract}
In this paper, we propose pretty simple
password-authenticated key-exchange protocol which is based on
the difficulty of solving DDH problem.
It has the following advantages: (1) 
Both $y_1$ and $y_2$ in our protocol
are independent and thus 
they can be pre-computed and can be sent independently.
This speeds up the protocol.
(2) Clients and servers can
 use almost the same algorithm.
This reduces the implementation costs without accepting replay
 attacks and abuse of entities as oracles.

{key words: password, key exchange, authentication}
\end{abstract}

\section{Introduction}

We consider the following situation.
Two entities, at least one of them is a human, beforehand
share a human 
memorable password, which is secure against on-line (and
serial) exhaustive searches, but vulnerable against
off-line (and parallel) exhaustive searches. Human entities 
 have only passwords in mind and have
no unmemorable secrets, such as private-keys, public-keys
(or fingerprints of them), secret information to use ID-based
cryptosystems.
Two entities run a protocol and share a new secret (we call it
keying material) that is
secure  
against off-line exhaustive searches.
The shared keying material is then used 
 to generate keys 
 for identifying the other entity and then  
 establishing a secure
 channel
(where secrecy and/or data integrity are provided).

While such secure channels can be established using
public-keys like SSH and SSL,  users must 
verify the validity of the public-keys used in them
 (using signature-verification keys or fingerprints of 
the public-keys).
For ordinal users, it is very troublesome to 
 carry them
 anywhere and anytime, and then perform verification.
Due to this troublesomeness, users may
skip the verification of the public-keys and weaken the
security of it. 

Password-authenticated key-exchanges are
 very convenient for users (especially when they log in
 their own servers remotely with their hands empty)
since they do not need to carry any 
 verification-keys or fingerprints with them and do
 not need to verify the public-keys for PKI \footnote{
One of the advantages of PKI is that unknown users can communicate securely.}.  
While such protocols have been proposed in 
\cite{GL01,MacK01b,KOY01,Jab01,MacK01a,Kw01,BMP00,BPR00},
most of them are a little bit complicated.

In this paper, we propose pretty-simple 
protocol which is based on the difficulty of solving DDH problem.
It has the following advantages: (1) 
Both $y_1$ and $y_2$ are independent and thus they
can be pre-computed and sent independently.
This speeds up the protocol without leaking the information on
the passwords.
(2) Clients and servers can
 use almost the same algorithm.
This reduces the implementation costs without accepting replay
 attacks and abuse of entities as oracles.

\section{Our Protocol}

Our protocol is defined over 
 a finite cyclic group ${\cal G} = <g>$ where
 $|{\cal G}|=q$ and $q$ is a large prime (or 
a positive integer divisible by a large prime).
While ${\cal G}$ can be a group over an elliptic curve, 
 in this paper we assume ${\cal G}$ is a prime order subgroup
 over a finite field $F_p$.

Both $g$ and $h$ are two generators of ${\cal G}$, chosen 
 so that
 its DLP
 (Discrete Logarithm Problem), i.e.
  calculating 
 \begin{eqnarray}
   a = \log_g h ,
 \end{eqnarray}
 should be hard\footnote{Since we assume
the DDH (Decision Diffie-Hellman) problem is hard, it is
reasonable to assume that DLP is also hard.} for each entity.
Both $g$ and $h$ may be chosen as system parameters
 or chosen with the negotiation between entities.
For example, 
$g$ may be a random generator of ${\cal G}$ and
$h:=Hash(g)^{(p-1)/q} \mod p$, or
 a client chooses $g:=g_b^{s_1}$ for a random $s_1 \in
\Zq$ where $g_b$ is a random generator of ${\cal G}$, and then
sends its commitment 
$Hash(g)$ to a server,  the server 
 replies $h:=g_b^{s_2}$ for a random $s_2 \in
\Zq$, and finally the client reveals $g$
to the server.

The protocol consists of 
the following two phases: 
a secrecy-amplification phase and a verification phase.

In the secrecy-amplification phase, the secrecy of a
pre-shared weak secret, i.e. a human memorable password that
may be vulnerable against off-line attack, is amplified to a
strong secret, i.e. a keying material that is secure even against off-line attack.
In the verification phase, an ordinal challenge-response
 protocol
 is used to verify whether the other entity has the
 same secret or not.
The point to notice is that 
challenges should be chosen to be unique at every
 session and at every entity,
and to be uncontrollable by an entity in one side
 to
 avoid replay attacks and abuse of one entity in the other
 side as an oracle.

Both phases are describe as follows.

\subsection{Secrecy-Amplification Phase}

\begin{figure*}[t]
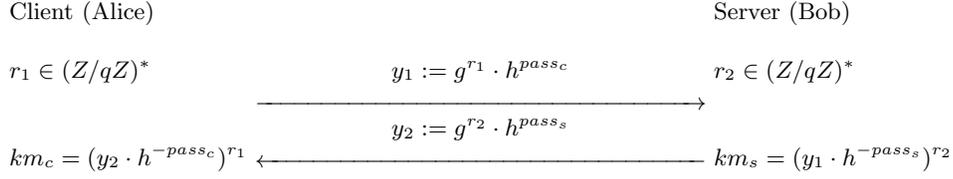

  \centering
  \begin{tabular}{lcl}
Client (Alice) & &  Server (Bob) \\
&& \\
$r_1 \in \Zq$ & $y_1:= g^{r_1} \cdot h^{pass_c}$  &   $r_2 \in \Zq$\\
 & $\overrightarrow{\hspace{\arrowsize}}$ & \\
& $y_2:= g^{r_2} \cdot h^{pass_s}$  &  \\
 $km_c = (y_2 \cdot h^{-pass_c})^{r_1}$ & $\overleftarrow{\hspace{\arrowsize}}$ &  $km_s = (y_1 \cdot h^{-pass_s})^{r_2}$ \\
  \end{tabular}
  \caption{Secrecy-amplification phase of our protocol}
  \label{fig:protocoli}
\end{figure*}

The secrecy-amplification phase
 is illustrated in Fig. \ref{fig:protocoli}.
%
A client
 chooses a random number $r_1 \in \Zq$ and
then calculates $y_1:= g^{r_1} \cdot h^{pass_c}$ using its
password $pass_c$. 
It sends $y_1$ to a server.
The server also  
 calculates $y_2:= g^{r_2} \cdot h^{pass_s}$ using its
password $pass_s$ and a random number $r_2 \in \Zq$, and
then sends it to the client
Now,  the client's keying
material is $km_c = (y_2 \cdot h^{-pass_c})^{r_1}$ and 
the server's 
one is $km_s = (y_1 \cdot h^{-pass_s})^{r_2}$. 

Only when they run the protocol using the same password, they
can share the same keying material. Otherwise 
distinguishing the other's one
 is as hard as solving DDH problem that
 is defined as follows:
\begin{deff}
({\bf DDH problem})
Given $g_b \in {\cal G}$ and
$d=(d_1,d_2,d_3)=(g_b^{x_1},g_b^{x_2},g_b^{x_3})$  
where $x_3$ is either  $x_1 x_2$ or not with probability $1/2$,
then decide whether $g_b^{x_3} = g_b^{x_1 x_2}$ or not.
\end{deff}

One of the  advantages of this protocol is 
 that both $y_1$ and $y_2$ are independent and thus they
can be pre-computed and sent independently.
This means the servers can transmit $y_2$ first (or before it
 receives $y_1$). 
This speeds up the protocol without leaking the information of
 the passwords since  
they are masked with random numbers $r_2$ (or
 $r_1$).

Another advantage is that both the clients and the servers can
 use almost the same algorithm.
This reduces the implementation costs without
accepting replay
 attacks and abuse of entities as oracles since 
 $(y_1,y_2)$ cannot be controlled by one entity and it is unique at every sessions and entities.

\subsection{Verification Phase}

Whether the other entity shares the same keying material
with me is verified in this phase as follows:
Both the client and the server exchange
\hspace{1.5em}
 $v_1 :=$  $KH_{km_s}(Tag_s||y_1||y_2)$ and
 $v_2 := KH_{km_c}(Tag_c||y_1||y_2)$ each other 
where $v_1$ is generated by the server and $v_2$  is generated
by the client respectively,
$KH_k()$ is a keyed hash
function whose key is $k$. Both $Tag_s$ and $Tag_c$ are
pre-determined distinct values, e.g. 
$Tag_s=0$ and $Tag_c=1$.   
The client verifies 
 $v_1 \stackrel{?}{=} KH_{km_c}(Tag_s||y_1||y_2)$
 and the server verifies
 $v_2 \stackrel{?}{=} KH_{km_s}(Tag_c||y_1||y_2)$.

Similarly to the secrecy-amplification phase,
both $v_1$ and $v_2$ can be transmitted independently each
other. (This verification phase may be skipped if
data-integrity is provided after the secrecy-amplification
phase using the shared keying
material.)

While adversaries can perform exhaustive searches for the
keying material using $v_1$ or $v_2$, that is not a matter if 
strong secret can be shared at the
 secrecy-amplification phase and 
no efficient algorithm is known to find the key $k$ of 
  $KH_k()$ than exhaustive searches.
The latter property can be satisfied using 
practical functions, such as 
 HMAC \cite{hmac} so far, and then 
 $KH_k()$ does not need to be a random oracle.

\section{Conclusion}

We proposed pretty simple 
password-authenticated key-exchange 
protocol which is base on the difficulty of solving the DDH
problem. 

Our protocol has the following advantages: (1) 
both $y_1$ and $y_2$ are independent and thus they
can be pre-computed and sent independently.
This speeds up the protocol, but
 does not leak the information on the passwords
since they are masked with random numbers $r_1$ (or $r_2$).
(2) Clients and servers can
 use almost the same algorithm.
This reduces the implementation costs, but
 does not weaken the security against replay attacks and  
 abuse of entities as oracles
 since $(y_1,y_2)$ cannot be controlled by one entity and it is unique at every sessions and entities.

\bibliographystyle{plain}
\bibliography{pub,secu}

\end{document}